# RADIATION AND THERMAL ANALYSIS OF PRODUCTION SOLENOID FOR MU2E EXPERIMENTAL SETUP[*,†]

V.S. Pronskikh[#], V.V. Kashikhin, N.V. Mokhov, Fermilab, Batavia, IL 60510, U.S.A.

## Abstract

The Muon-to-Electron (Mu2e) experiment at Fermilab, will seek the evidence of direct muon to electron conversion at the sensitivity level where it cannot be explained by the Standard Model. An 8-GeV 25-kW proton beam will be directed onto a tilted gold target inside a large-bore superconducting Production Solenoid (PS) with the peak field on the axis of ~5T. The negative muons resulting from the pion decay will be captured in the PS aperture and directed by an S-shaped Transport Solenoid towards the stopping target inside the Detector Solenoid. In order for the superconducting magnets to operate reliably and with a sufficient safety margin, the peak neutron flux entering the coils must be reduced by 3 orders of magnitude that is achieved by means of a sophisticated absorber placed in the magnet aperture. The proposed absorber, consisting of W- and Cu-based alloy parts, is optimized for the performance and cost. Results of MARS15 simulations of energy deposition and radiation are reported. The results of the PS magnet thermal analysis, coordinated with the coil cooling scheme, are reported as well for the selected absorber design.

[*]Work supported by Fermi Research Alliance, LLC under contract No. DE-AC02-07CH11359 with the U.S. Department of Energy.
[†]Presented paper at Particle Accelerator Conference'11, March 28 – April 1, 2011, New York, U.S.A.
[#]vspron@fnal.gov

# RADIATION AND THERMAL ANALYSIS OF PRODUCTION SOLENOID FOR MU2E EXPERIMENTAL SETUP*

V.S. Pronskikh#, V.V. Kashikhin, N.V. Mokhov, FNAL, Batavia, IL 60510, U.S.A.

*Abstract*

   The Muon-to-Electron (Mu2e) [1] experiment at Fermilab, will seek the evidence of direct muon to electron conversion at the sensitivity level where it cannot be explained by the Standard Model. An 8-GeV 25-kW proton beam will be directed onto a tilted gold target inside a large-bore superconducting Production Solenoid (PS) with the peak field on the axis of ~5T. The negative muons resulting from the pion decay will be captured in the PS aperture and directed by an S-shaped Transport Solenoid towards the stopping target inside the Detector Solenoid. In order for the superconducting magnets to operate reliably and with a sufficient safety margin, the peak neutron flux entering the coils must be reduced by 3 orders of magnitude that is achieved by means of a sophisticated absorber placed in the magnet aperture. The proposed absorber, consisting of W- and Cu-based alloy parts, is optimized for the performance and cost. Results of MARS15 [2] simulations of energy deposition and radiation are reported. The results of the PS magnet thermal analysis, coordinated with the coil cooling scheme, are reported as well for the selected absorber design.

## INTRODUCTION

   The Mu2e experiment will be devoted to studies of the charged lepton flavor violation (CLFV) which up to now has never been observed and can manifest itself as the conversion of $\mu^-$ to $e^-$ in the field of a nucleus without emission of neutrinos. The emission of monoenergetic 105-MeV electrons can serve as a signature of such a process. CLFV has a very low probability ($<10^{-54}$) in the Standard Model, and its observation by the Mu2e [1] experiment can find its explanation in SUSY theories, extra dimensions, leptoquarks, compositeness, second Higgs doublet etc. Mu2e experiment is a further development of the proposed MECO and MELC experiments.

   One of the main parts of the Mu2e experimental setup is its production solenoid, in which negative pions are generated in interactions of the primary proton beam with the target (see Fig. 1). These pions then decay into muons which are delivered by the transport solenoid to the detectors. The off-axis 8 GeV proton beam will deliver $2 \cdot 10^{13}$ protons per second ($3 \cdot 10^7$ protons per pulse, every 1.7 μs) to the gold target, placed at the center of the PS bore.

## PS ABSORBER OPTIMIZATION

   The constraints in the PS absorber design are quench stability of the superconducting coils, low dynamic heat loads to the cryogenic system, a reasonable lifetime of the coil components, acceptable hands-on maintenance conditions, compactness of the absorber that should fit into the PS bore and provide an aperture large enough to not compromise pion collection efficiency, cost, weight and other engineering constraints. The following ten versions have been studied in the course of PS absorber material optimization:

- entirely tungsten absorber, denoted #1 in Tables
- five multilayer absorbers (Figure 1), composed of
  a. 5cm W, 20 cm Fe, 12 cm $BCH_2$, 3 cm Fe (from inside outside), #2
  b. 5cm W, 20 cm Fe, 12 cm $BCH_2$, 3 cm Cd, #3
  c. 5cm W, 20 cm Fe, 12 cm $BCH_2$, 3 cm Cu, #4
  d. 5cm W, 20 cm Cu, 12 cm $BCH_2$, 3 cm Fe, #5
  e. 5cm W, 20 cm WC, 12 cm $BCH_2$, 3 cm Fe, #6
- entirely tungsten carbide (WC) absorber, #7
- entirely depleted $^{238}$U absorber, #8
- tungsten absorber with a conic copper part, #9
- tungsten absorber with a cylindrical copper part (Figure 2), #10.

   Tungsten carbide (WC) means a mix of 80% WC and 20% water. The composition of the multilayer absorbers was chosen according to their purposes: 1-st layer to stop charged particles, 2-nd layer to slow down the neutrons resulting from interactions in the target and 1-st layer, 3-rd layer to capture the slow neutrons and 4-th layer to suppress γ-quanta resulting from neutron capture.

   When optimizing the absorber, the following parameters were taken into account: dynamic heat load, peak power density, number of displacements per atom (DPA) in the helium-cooled solenoid coils, peak prompt dose and peak neutron flux in the superconducting coils [3]. As one of the primary functions of the absorber is to protect the coils from warming and consequential quench, first two parameters serve to determine if the critical heating is attained and also to determine requirements to the cooling system. Details of the magnet design are reported in [4].

   The highest power densities were attained in the most absorbers in the first coil (only in #10 in the second), then in the second (#10 in the first), and the lowest (one or two orders of magnitude less) in the third. In all the considered absorbers except for the multilayer ones the power density values are not critical from the requirement point of view, however, the most promising from this point of view are absorbers #9 and #10.

   Absorber #10 was taken as a baseline for further optimization (Figure 3): made from bronze (which has better magnetic properties and is thus safer from the point of view of quench protection) and a W-alloy, which has better mechanical properties then pure tungsten. Also, the downstream (in Figures – on the left from the target) part of the absorber is one third from a W-alloy, and two thirds from bronze, so that the parts nearest to the spent beam exiting the absorber were from W-alloy, while the others – from bronze. This approach allowed reducing the necessary amount of tungsten significantly.

___

*Work supported by Fermi Research Alliance, LLC under contract No. DE-AC02-07CH11359 with the U.S. Department of Energy.
#vspron@fnal.gov

Figures 4-5 show dynamic heat load and DPA in the coils with the optimised absorber, the values satisfy the requirements [3, 5] and thus the absorber (see Figure 3) was chosen as the baseline.

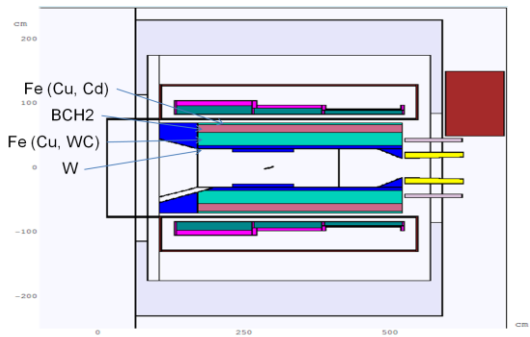

Figure 1. Cross section of a multilayer absorber (#2-#6).

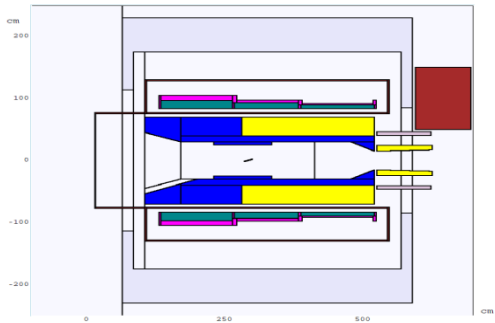

Figure 2. Cross section of a tungsten absorber with a cylindrical copper part.

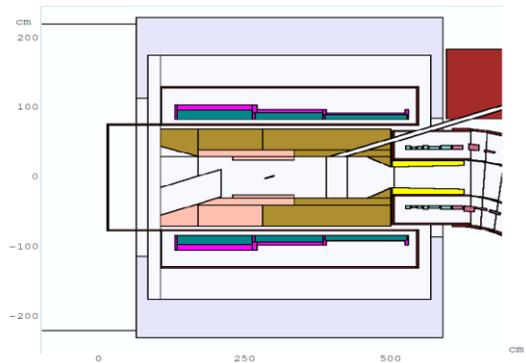

Figure 3. Cross section of the optimised absorber of W-alloy and bronze.

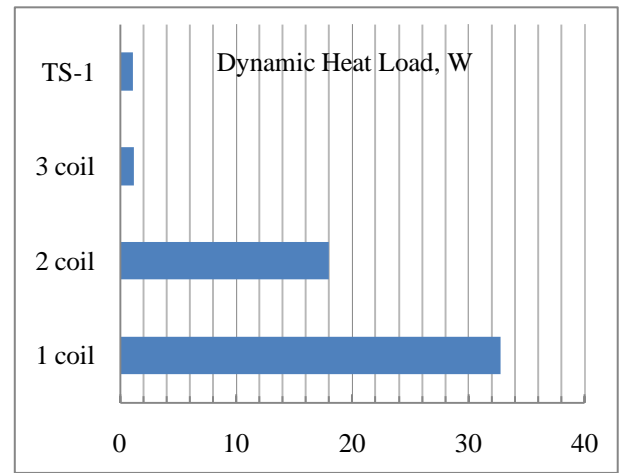

Figure 4. Dynamic heatload in PS coils with the optimised absorber, W

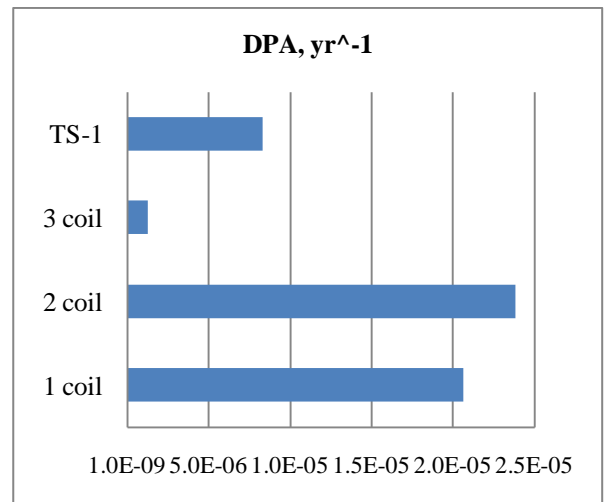

Figure 5. DPA in PS coils with the optimized absorber, $yr^{-1}$

## THERMAL ANALYSIS

The 3D thermal analysis was performed for the radiation heat load in case of the optimized absorber design presented earlier. The FEM model created by COMSOL Multiphysics was discretized to the level of individual layers and the interlayer insulation/conducting sheets. The thermal conductivity of each layer in the axial direction was modeled by the equivalent thermal conductivity of the insulated cable in that direction.

In the radial and azimuthal direction, the coil layers were assigned the actual properties of Al and Cu, with the appropriate reduction of thermal conductivities due to the irradiation. The radial cable insulation was combined with the interlayer thermal insulation; both assigned the properties of G10. The ground insulation between the coil and the support structure included 2 mm of G10.

The coil layers were separated from each other by two layers of insulation with a layer of Al in between that was 1-2mm thick, depending on the location within the coil. The Al layers formed thermal bridges by connecting to the Al plates placed between the coil ends and the end flanges. These plates extended all the way to the outer cold mass surface, connecting there to the cooling tubes. The Al layers and plates formed the thermal bridges connecting coils to the cooling system.

Figure 6 shows the 3D thermal model of the PS cold mass with the cooling tubes attached to the outer surfaces of the support shells. The dynamic heat load map in the coil and the support structure generated by the MARS code was applied to all parts of the cold mass. In addition to that, the relevant static heat loads were applied to all external surfaces to model the thermal radiation/gas conduction; to the middle support ring to model the heat load through the axial support; and to the end flanges to model the heat load through the transverse supports. It was assumed that the cooling tubes are kept at the constant temperature $T_0$ by the cryogenic system.

The resulting temperature in the cold mass is shown in Figure 7 for $T_0 = 4.5$ K. The maximum temperature is in the middle of the inner surface of the thickest coil section; that location coincides with the peak field location, and, therefore directly affects the thermal margin.

In order to determine the thermal parameter space, $T_0$ was varied in the 4.2 K – 4.8 K range. Figure 8 shows the peak coil temperature as a function of $T_0$ for static and static+dynamic heat loads. The $T_0$ temperature can be as high as 4.7 K with the static-only heat load to meet the operating requirement for the temperature margin of 1.5 K. In order for $T_0$ to meet the same temperature margin at the nominal static+dynamic heat load, $T_0$ must not exceed 4.48 K that is achievable by the foreseen cryogenic system.

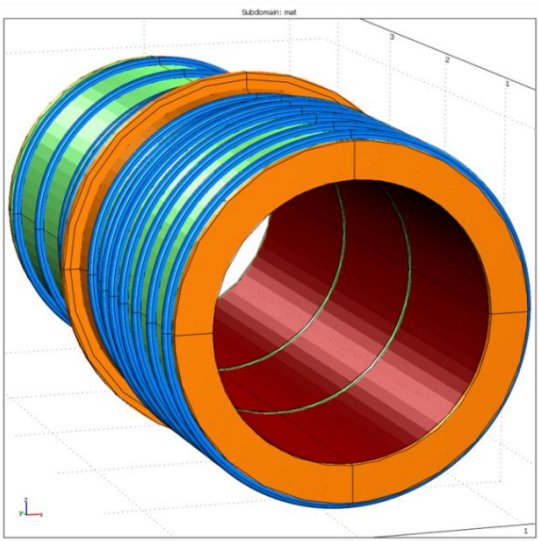

Figure 6. 3D model of PS cold mass.

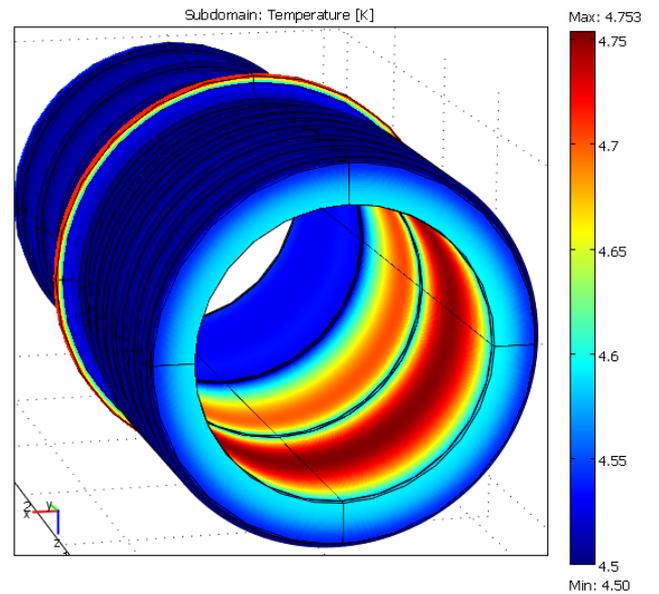

Figure 7. Temperature distribution in the cold mass under static and dynamic heat loads for $T_0$=4.5 K.

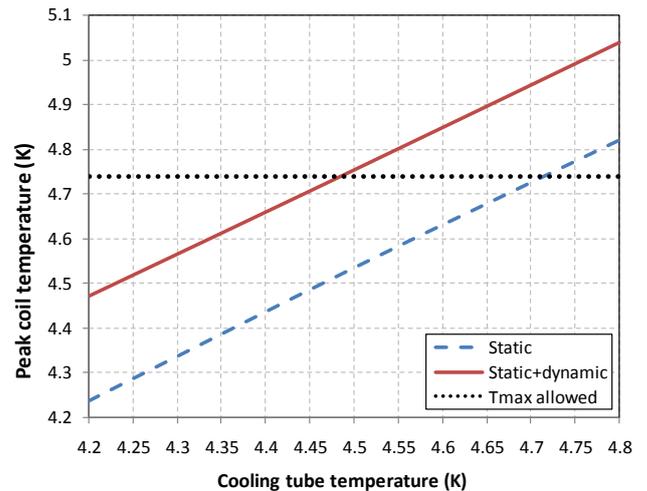

Figure 8. Peak coil temperature as function of $T_0$.

## REFERENCES

[1] http://mu2e.fnal.gov
[2] http://www-ap.fnal.gov/MARS/
[3] J. Popp, R. Coleman, V. Pronskikh, Requirements for the Mu2e Production Solenoid Heat and Radiation Shield, Mu2e-doc-1092.
[4] G. Ambrosio at al., Design studies of Mu2e Production Solenoid, Mu2e-doc-1110.
[5] V.S. Pronskikh, N.V. Mokhov, Proceedings of XX International Baldin Seminar on High Energy Physics Problems, Dubna, October 4-9, 2010, Fermilab-Conf-11-024-APC.